\documentclass[pre,twocolumn,showpacs,amsmath,amssymb,showkeys,nofootinbib,floatfix]{revtex4}

\usepackage{graphicx}
\usepackage{dcolumn}
\usepackage{bm}
\usepackage{times}

\begin{document}


\title{Dynamics of bimodality in vehicular traffic flows}

\author{Arjun Mullick}
\email{arjunmullick@gmail.com}
\altaffiliation{Department of Computer Science Engineering.
(Present address: {\it Infosys Limited}, Plot No. 24, Rajiv 
Gandhi Infotech Park, Phase--II, Village--Man, Taluka--Mulshi,
Pune 411057, India. E-mail: {\tt arjun\_mullick@infosys.com})}
\author{Arnab K. Ray}
\email{arnab.kumar@juet.ac.in}
\altaffiliation{Department of Physics}
\affiliation{Jaypee University of Engineering 
and Technology, 
Raghogarh, Guna 473226, Madhya Pradesh, India}

\begin{abstract}
A model equation has been proposed to describe bimodal features
in vehicular traffic flows. The dynamics of the bimodal distribution
reveals the existence of a fixed point that is connected to itself 
by a homoclinic trajectory. The mathematical conditions associated 
with bimodality have been established. The critical 
factors necessary for both a breaking of symmetry and a transition 
from bimodal to unimodal behaviour, in the manner
of a bifurcation, have been analysed.
\end{abstract}

\maketitle

\section{Introduction}
\label{sec1}

Bimodal distributions are characterised by the existence 
of two distinct modes (peaks) in a single frequency 
distribution~\cite{ross}. Usually such a distribution is 
seen as a mixture of two normal distributions.  
Standard methods to understand bimodal distributions
take recourse to statistical methods, and it is only 
rarely that a model of bimodality is 
provided~\cite{daorblo}. 
In keeping with the latter principle, the primary emphasis 
of this work is on developing a mathematical model that will
give a single global description of bimodal phenomena. 
The model proposed here has been structured on the bimodal
distribution of data pertaining to vehicular traffic flows,
which is a man-made condition, although bimodal features 
are exhibited in many natural phenomena as
well~\cite{guthrie,chhe,covita}. From a purely 
mathematical perspective, however, the scope of the model 
goes much beyond any particular practical context. 

The model has been used to understand the 
dynamics underlying a bimodal distribution, following 
standard methods of nonlinear dynamics~\cite{stro,js}. 
In so doing, the dependence of bimodal properties on 
various parameters has been brought to the fore. The 
dynamics of the model bimodal distribution reveals that 
the controlling parameters can be  adjusted to cause 
bimodal-to-unimodal transition (like a bifurcation) 
and a breaking of symmetry in the system (both of 
which have been seen in some previous studies 
on bimodality~\cite{bibile,dyguno}). 

\section{Bimodality in Vehicular Traffic Flows}
\label{sec2}

Nowadays   
the number of vehicles on the roads has increased prodigiously, 
along with an accompanying necessity to expand road networks.
Any implementation of an advanced transportation system, 
therefore, needs a clear view of the present intricacies in 
traffic flows. Studies of vehicular traffic flows
usually aim at understanding interactions among streams of 
vehicles (modelled occasionally as particles), 
and a substantial body of relevant work, ranging across multiple
perspectives, has already come into 
existence~\cite{chosas,helbing,wald1,wald2,riolar,kamaha,maemoor1,
hejitre, maemoor2,bkc,refra,car,ger,chavamoba,prahac,rohis,gerros,
gigaro,nekepar,dingding}. The practical motive
behind these studies is to eliminate the problem
of traffic congestion, and to devise effective methods of 
controlling traffic flow. 

The mathematical model in this work
takes a macroscopic view of the vehicular traffic of a city
at a given hour in the day --- in effect, the bulk of traffic
as a function of time. The data needed for the purpose of 
modelling have been taken from the repository of the Alabama 
Department of Transportation 
(ALDOT)~\footnote{\tt{http://aldotgis.dot.state.al.us/atd/default.aspx}}, 
and the particular city whose traffic data have been used here, 
is Jackson in Alabama State, USA. The principal factor guiding
the choice of this city is that with Jackson being comparatively 
small in size, the data about its traffic have been 
systematically recorded, and are free of the usual 
complications associated with the traffic
of very large cities. This has the advantage of simplicity, where 
placing the foundations of a mathematical model is concerned.   

The Jackson data indicate that the bulk traffic is dominated 
by flows in the eastward and westward directions. The plot of 
the data for the westward traffic is shown in Fig.~\ref{f1},
with the broken line. Similarly, the plot of the eastward traffic
is shown in Fig.~\ref{f2}. In both these plots, the time, $t$, 
at which the traffic volume, $N\left(t\right)$, was measured,
has been scaled so that $t=0$ coincides with midday, 
$12$:$00$ hours. This enables one to take advantage of any 
possible symmetry in the distribution of the data, especially
if there is a symmetry about $t=0$. From a look at 
both Figs.~\ref{f1}~\&~\ref{f2}, however, 
only an asymmetric bimodal distribution is to be seen, with two
peaks of unequal height in either plot. 
What Fig.~\ref{f1} suggests is that the morning traffic volume 
is much higher in the western direction than in the opposite
direction, while the behaviour of the bimodal curve in 
Fig.~\ref{f2} is quite the opposite, with the eastward traffic
being disproportionately high in the evening. There is 
a clear commuter preference in the traffic flow as a function
of time.

Extraneous traffic (due to vehicles coming from outside the
city, passing through it, and leaving it ultimately) may add 
to the count of the indigenous traffic flow. 
One may imagine Jackson city as a bounded volume,
through which there occurs a bulk flow of traffic from the 
rest of the world. This flow is bi-directional, and along 
either direction the 
external traffic volume maintains a uniform
average value. Besides, whatever goes into this 
bounded system at one end, also comes out of the 
other end in finite time. Consequently, the alien traffic 
does not cause any appreciable bias in the bimodal distribution. 
The variation in the data around this constant background of 
global traffic, is then due only to the local city traffic,
which oscillates back and forth between the eastern and western
boundaries, once in a cycle of twenty four hours. 

Social and economic conditions also have a major role to 
play in the way the data are 
distributed from one day of the week to the other. 
The weekend traffic (on Saturdays and 
Sundays) will follow a different pattern, in comparison to 
the traffic on a working day. And even on working days close
to the weekend, like Mondays and Fridays, the traffic flow may
be different from what it would be in the middle of the week. 
So a judicious choice of the data has to be made here. For the
purpose of crafting a mathematical model from first principles,
the data employed in this work are the average values of all
the Wednesday traffic data of Jackson city in the year, $2007$. 
Wednesday is in the middle of the week, and it is in this period
of the week that all city traffic is expected to rid 
itself of the weekend effect and reach a local mid-week 
equilibrium. So the corresponding data should convey more 
unambiguous trends in the traffic flow. 

\section{A Mathematical Model for Bimodality}
\label{sec3}

\begin{figure}[floatfix]
\begin{center}
\includegraphics[scale=0.70, angle=0]{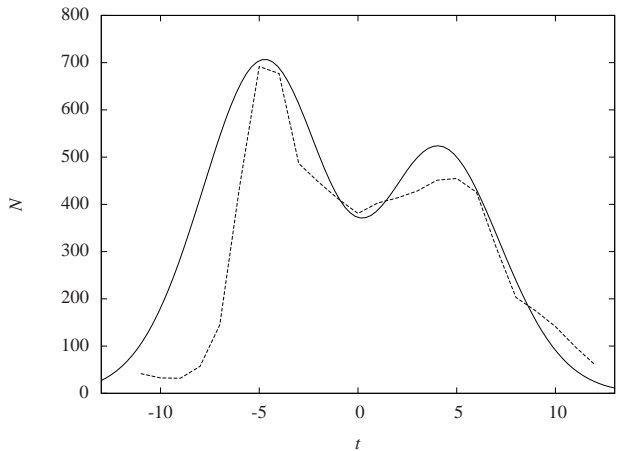}
\caption{\label{f1}\small{Bimodal distribution of traffic flow
due west. The vertical axis measures the volume of traffic, $N$,
for a given time, $t$ (scaled in hours). The horizontal axis 
has been scaled so that $t=0$ is set at midday, $12$:$00$ hours. 
The dotted broken curve joins the real data points, and the
continuous curve traces the
model given by Eq.~(\ref{modeq}).
The parameter values used for the fit are $A=44.0$, $\mu=8.53$,
$\lambda =0.19$ and $\beta = -0.09$.
}}
\end{center}
\end{figure}
\begin{figure}[floatfix]
\begin{center}
\includegraphics[scale=0.70, angle=0]{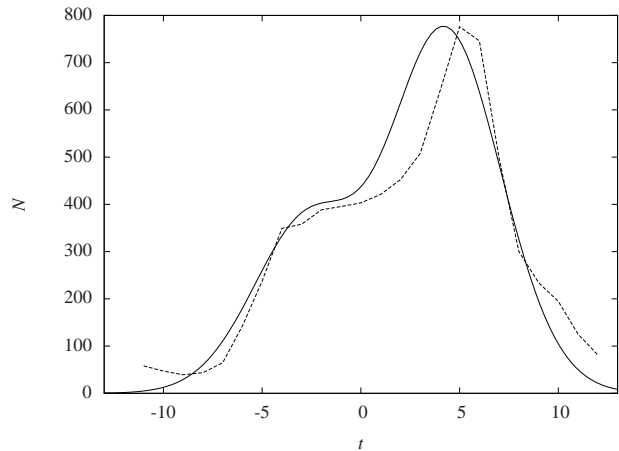}
\caption{\label{f2}\small{Bimodal distribution of traffic flow due
east. The plotting follows the same principles outlined in Fig.~\ref{f1},
but the parameter values necessary for fitting the data are
$A=44.1$, $\mu=10.5$, $\lambda=0.22$ and $\beta=0.24$. The sign
and the absolute
magnitude of $\beta$ make the most significant difference
between the two bimodal distributions.
}}
\end{center}
\end{figure}
The continuous curves in Figs.~\ref{f1}~\&~\ref{f2} follow the 
model function that describes bimodal distributions. 
In such distributions one encounters three turning points. So 
it is natural to suggest that a fourth-degree (quartic)
polynomial~\cite{bibile,dyguno},  
whose derivative is of the third degree (cubic), 
will correspond to the three turning points. 
This particular feature can be captured 
by an ``inverted Mexican hat"-type function of the form,
$N\left(t\right) = a + t^2 -bt^4$,
in which $N$ is volume of the traffic flow and $t$ is the 
{\em hour} at which the traffic flow has been measured. 
The discrepancy between this function and the pattern 
followed by the data, however, lies in the 
asymptotic conditions of the latter. Both the inner and the
outer time limits on the horizontal axis 
in Figs.~\ref{f1}~\&~\ref{f2}, are close to midnight. At that
hour, on a working day in the middle of the week, the volume 
of traffic trickles down to small values, a condition that 
should be described mathematically by a function with an
asymptotic decay, $N \longrightarrow 0$, at the extreme ends
of the time axis. And this is how it should be, because 
a bimodal distribution is usually treated as a 
mixture of two normal distributions, which have a similarly
decaying tail. This, however, is not indicated by the 
``inverted Mexican hat" function, which 
intersects the line $N=0$ in finite time. 

To capture the required asymptotic behaviour of the bimodal
distribution, therefore, it is expedient to employ a 
Maxwell-Boltzmann type of function in the form,
$N\left(t\right)=\alpha \left(\mu+t^2\right) 
\exp\left(-\lambda^2 t^2\right)$.
From the foregoing function, for low values of $t$, on 
approximating
$\exp\left(-\lambda^2 t^2\right)\simeq 1-\lambda^2 t^2$, 
one recovers the features of the ``inverted
Mexican hat" function. Another aspect of the Maxwell-Boltzmann
distribution is that its integral (the area under the curve)
is constant, in consonance with the fact 
that in an average sense the total volume of the internal city 
traffic is constant over one full day. 

With the required asymptotic condition satisfied thus, 
the role of the parameter, $\mu$, can now be examined. 
From the Maxwell-Boltzmann function, 
setting its time derivative, $\dot{N}=0$,
one obtains the condition, 
$t\left[1-\lambda^2\left(\mu+t^2\right)\right]=0$, which 
indicates three turning points in time, at $t=0$ and 
$t=\pm\left(\lambda^{-2}-\mu\right)^{1/2}$. Now, the two non-zero
roots of $t$ can only be real when $\mu < \lambda^{-2}$. When 
$\mu = \lambda^{-2}$, all the three roots coalesce at $t=0$. 
Noting that bimodality implies the existence of three turning
points, the importance of the parameter
$\mu$ in maintaining bimodality is now evident. 

The distribution traced out by the Maxwell-Boltzmann function
is, however, symmetric about $t=0$, which is certainly not to be
expected in a real bimodal distribution. And more to the point
for traffic flows, this lack of symmetry is exchanged 
over a time scale of half a day, as it has been shown in 
Figs.~\ref{f1}~\&~\ref{f2}. While the traffic due west is higher
in the morning hour than at any other time of the day, the traffic
due east displays an exact reversal of this trend. To account for
these two features, two new parameters, $\gamma$ and $\sigma$, 
are now introduced to break the symmetry of the 
Maxwell-Boltzmann distribution, which, consequently, goes as
\begin{equation}
\label{mbasym}
N\left(t\right)=\alpha \left(\mu +t^2\right) 
\exp\left(-\lambda^2 t^2\right)
\left(1+\gamma \right)^{\sigma t}\,.
\end{equation}
Any non-zero value of $\gamma$ in the foregoing expression will 
raise one peak in the bimodal distribution, and lower the other,
i.e. break the symmetry about $t=0$ in Figs.~\ref{f1}~\&~\ref{f2}. 
Symmetry is, of course, restored when $\gamma =0$, and in this 
special case, Eq.~(\ref{mbasym}) 
assumes the general mathematical 
form of the energy eigenfunction of the second 
state of the linear harmonic oscillator~\cite{schiff}. When 
$\gamma \neq 0$, the part played by $\sigma$ in the
distribution given by Eq.~(\ref{mbasym}) is to exchange the 
respective 
positions of the two uneven peaks, depending on the direction 
of the traffic flow, a condition that can be achieved simply by 
assigning the values, $\sigma =\pm 1$. 

With all the parameter values specified suitably, Eq.~(\ref{mbasym})
can be used to fit the data, but a transformation of it
helps in reducing the number of parameters in the model. 
Accordingly, Eq.~(\ref{mbasym}) is recast as 
\begin{equation}
\label{modeq}
N(t)= A \left(\mu +t^2\right) 
\exp\left[-\left(\lambda t-\beta\right)^2\right]\,,
\end{equation}
in which 
$\beta =\sigma \left(2\lambda\right)^{-1}
\ln\left(1+\gamma\right)$ and $A=\alpha \exp\left(\beta^2\right)$. 
Now $\beta$, having taken over the roles of 
$\gamma$ and $\sigma$, controls both the breaking of symmetry and the
exchange of the peaks in the bimodal curve. The former condition is 
obtained from the 
absolute magnitude of $\beta$, while the latter from its
sign. These values, as well as values of $A$, $\mu$ and $\lambda$
are used to fit the data, as shown 
in Figs.~\ref{f1}~\&~\ref{f2}. 

\section{Dynamics of Bimodality}
\label{sec4}

The model for the data fit, as given by Eq.~(\ref{modeq}), is in 
the form, $N\equiv N\left(t\right)$. Two parameters in this model, 
$\mu$ and $\beta$, play a crucial part in causing an asymmetric 
bimodality. For certain values of these parameters (for instance 
$\mu = \lambda^{-2}$ and $\beta =0$), the bimodal distribution 
undergoes qualitative changes. To investigate these aspects of 
the model in detail, it will be  
necessary to find $\dot{N}$, and then to plot
a phase portrait, $N$--$\dot{N}$. This will reveal much 
about the stability, creation and annihilation of 
the turning points, and also the properties of any possible 
fixed point(s) in the bimodal system~\cite{stro,js}. 
So going back to Eq.~(\ref{modeq}), its first derivative 
in time (indicated by an overdot) is obtained in a 
non-autonomous form as
\begin{equation}
\label{tempo}
\dot{N} = 2 \left[\frac{t}{\mu +t^2} 
-\lambda\left(\lambda t-\beta\right)\right]N\,.
\end{equation} 
Turning points are given by the condition $\dot{N}=0$. 
In addition to this requirement, fixed points are obtained 
when $\ddot{N}=0$~\cite{js}. So to know the conditions 
needed for determing the fixed points of the system, the
second time derivative of the function, $N\equiv N\left(t\right)$, 
is obtained and set down in a non-autonomous form as 
\begin{equation}
\label{deriv2}
\ddot{N}=2 \left[\frac{1-4\lambda t\left(\lambda t-\beta \right)}
{\mu + t^2} + 2\lambda^2\left(\lambda t-\beta\right)^2 
-\lambda^2\right]N\,. 
\end{equation} 
Both Eqs.~(\ref{tempo})~and~(\ref{deriv2}) indicate the existence 
of a fixed point at $N=N^\star =0$, when the conditions 
$\ddot{N}=\dot{N}=0$ are simultaneously satisfied. This fixed
point is obtained, going by Eq.~(\ref{modeq}), when 
$t\longrightarrow \pm \infty$. The existence of other fixed 
points will depend on whether or not 
Eqs.~(\ref{tempo})~and~(\ref{deriv2}) will have common roots 
in $t$ when $\ddot{N}=\dot{N}=0$. Solving for $t$ under these
conditions will show how $\mu$ and 
$\beta$ influence the dynamics of the bimodal system. 

The role that $\mu$ plays can be known clearly by considering
the limiting case of $\beta =0$. In this perfectly
symmetric limit, on solving the explicitly time-dependent 
part of $\dot{N}=0$, three roots 
of $t$ are obtained, one at $t=0$ and the other two at 
$t=\pm\left(\lambda^{-2}-\mu\right)^{1/2}$. None of these 
three roots of $t$ leads to $\ddot{N}=0$. The only conclusion
that can be drawn from this fact is that there is only one 
fixed point in the bimodal system, at $N=N^\star =0$.
All other solutions of $\dot{N}=0$ give only turning points 
in the bimodal system. For $t=0$ one obtains a turning point 
of $N$ at $N_0=A\mu$, as it can be seen from Eq.~(\ref{modeq}). 
The two non-zero roots of $t$ at 
$t=\pm\left(\lambda^{-2}-\mu\right)^{1/2}$, will give two 
coinciding turning points on the line $\dot{N}=0$, in the 
phase portrait. From Eq.~(\ref{modeq}), these two turning 
points are to be found at 
$N_{\mathrm T}=A\lambda^{-2}\exp \left(\mu \lambda^2-1\right)$. 
To have
three real roots of $t$, it is necessary to fulfil the condition, 
$\mu < \lambda^{-2}$. By means of this condition it can also be
argued that $N_{\mathrm T} > N_0$. Hence, as long as 
$\mu < \lambda^{-2}$, there shall remain three turning points, 
one at $N=N_0$ and the other two coinciding at $N=N_{\mathrm T}$. 
As $\mu \longrightarrow \lambda^{-2}$,
these two turning point positions approach each other 
in the $N$--$\dot{N}$ phase portrait. When $\mu = \lambda^{-2}$, 
there is a bifurcation-like behaviour because of which the 
turning point at $N_0$ and one of the turning points at 
$N_{\mathrm T}$ annihilate each other. What survives this 
mutual annihilation is the second ``concealed" turning point
at $N_{\mathrm T}$. Subsequently, for $\mu > \lambda^{-2}$, 
only this turning point will continue its existence, and the
distribution will become unimodal. 

\begin{figure}[floatfix]
\begin{center}
\includegraphics[scale=0.70, angle=0]{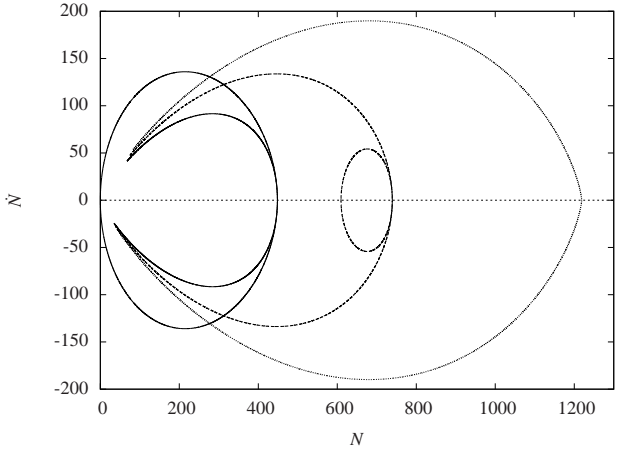}
\caption{\label{f3}\small{
A symmetric phase portrait (with $\beta =0$), using the
parameter values of $A$ and $\lambda$ taken from Fig.~\ref{f1}.
When $\mu =0$, the phase trajectories are shown by the continuous
curve at the extreme left. The closed loop has the largest area
in this case. When $\mu = 0.5\lambda^{-2}$, the loop size shrinks,
as the dotted curve shows in the centre of the plot.
When $\mu = \lambda^{-2}$, the loop vanishes as the curve on
the extreme right shows. The existence of a closed loop in the
plots above is a signature of bimodality. When $\mu = \lambda^{-2}$,
a bimodal-to-unimodal transition occurs, and the closed loop
disappears, i.e. bimodality is lost. All along there is a fixed 
point at 
$N=\dot{N}=0$, joined to itself by a homoclinic trajectory.}}
\end{center}
\end{figure}

The entire sequence of
this behaviour of the turning points over the range, 
$0 < \mu < \lambda^{-2}$, has been depicted in Fig.~\ref{f3}. 
It is evident from this plot that bimodality is closely
related to the existence of a closed loop for a phase trajectory 
in the $N$--$\dot{N}$ portrait. This is one clear message that 
has emerged from the phase-portrait analysis. 
When $\mu \longrightarrow 0$, 
bimodality becomes very pronounced, and when 
$\mu \longrightarrow \lambda^{-2}$,
bimodality becomes progressively enfeebled, disappearing 
altogether when $\mu = \lambda^{-2}$.  
In the former case, the sole fixed point at $N=N^\star=0$ 
coincides with the turning point, $N_0=0$, while in the latter 
case, bimodality is eradicated because of a mutual annihilation 
between the turning points, $N_0\left(\neq 0\right)$ and 
$N_{\mathrm T}$. Once bimodality disappears and the distribution
becomes unimodal, another fact that stands out clearly 
through the phase-portrait analysis is the existence of 
a homoclinic trajectory that joins the fixed point, 
$N^\star =0$, to itself. Also, the closed
loop that traces bimodality in the phase portrait  
indicates a local periodic behaviour (which effectively
implies that the two peaks in the bimodal distribution give
a local scale), and when $\mu \longrightarrow 0$, this local 
periodicity of what is now a strongly bimodal system, takes 
place on a time scale of $\lambda^{-1}$. 
\begin{figure}[floatfix]
\begin{center}
\includegraphics[scale=0.70, angle=0]{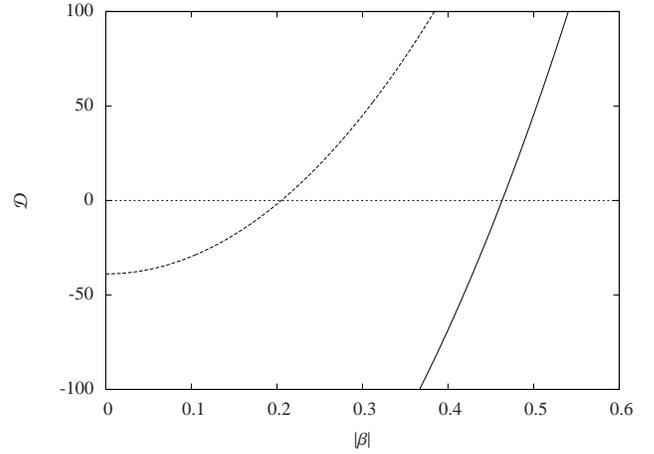}
\caption{\label{f4}\small{
When symmetry is broken (for $\beta \neq 0$), there are three
distinct turning points for $\dot{N}=0$. When the asymmetry
is large, two of these fixed points are lost due to a
bimodal-to-unimodal transition. This happens when the discriminant
of Eq.~(\ref{canon3}),
$\mathcal{D} =0$. The variation of $\mathcal{D}$ against
$\vert \beta \vert$ (since $\beta <0$ for the westward traffic)
shows that the transition occurs when
$\vert \beta \vert = \vert \beta_{\mathrm b} \vert \simeq 0.21$
for the eastward traffic (traced by the dotted curve), and when
$\vert \beta \vert = \vert \beta_{\mathrm b} \vert \simeq 0.46$
for the westward traffic (the continuous curve). For both of these
cases the values of $\mu$ and $\lambda$ are what they are for
Figs.~\ref{f1}~\&~\ref{f2}, respectively.}}
\end{center}
\end{figure}

The next effect to consider is the one of breaking the symmetry 
corresponding to $\beta =0$. 
The first thing to happen when $\beta \neq 0$, is that 
the two hitherto coinciding turning 
points at $N=N_{\mathrm T}$ are separated from each
other. A simple perturbative argument suffices to establish this. 
If $\beta \ll 1$, then about the turning point $N_{\mathrm T}$ 
(determined for $t^2=\lambda^{-2}-\mu$), it is possible to 
linearise all terms involving $\beta$ in Eq.~(\ref{modeq}), 
and obtain the two separated roots as 
$N_{{\mathrm T}\pm} \simeq N_{\mathrm T} 
\left[1 \pm 2\beta\left(1-\mu\lambda^2\right)^{1/2}\right]$. 

As $\beta$ increases in magnitude, the position 
of $N_{{\mathrm T}+}$ 
shifts to higher values of $N$ on the line $\dot{N}=0$, while the
position of $N_{{\mathrm T}-}$ slides in the opposite direction. 
This pair of roots corresponds to the two asymmetric peaks in the 
bimodal curve, with $N_{{\mathrm T}+}$ tracing the locus of the 
higher peak, and $N_{{\mathrm T}-}$ the lower. Between the two 
of them, these peaks flank a point of local minimum in the bimodal 
distribution, which itself turns out to be the third turning point 
on the line, $\dot{N}=0$. When $\beta =0$, this point is at $N=N_0$, 
with $N_0 < N_{\mathrm T}$, but when $\beta \neq 0$, the position 
of this third unpaired turning point shifts to larger values of 
$N$, and for sufficiently high values of $\beta$, it suffers a 
separate pair annihilation with the shifting turning point at 
$N_{{\mathrm T}-}$. This reasoning is borne out by solving for 
$\dot{N}=0$ in Eq.~(\ref{tempo}), specifically its
non-autonomous, time-dependent part. What transpires is a 
third-order polynomial in the variable $t$, which goes as    
\begin{equation}
\label{poly3}
t^3 -\left(\frac{\beta}{\lambda}\right)t^2 -
\left(\frac{1}{\lambda^2}-\mu \right)t-\frac{\mu \beta}{\lambda}=0\,.
\end{equation}
Performing a variable transformation, $t=\tau +h$, in the 
preceding relation, and setting 
$h=\beta \left(3\lambda\right)^{-1}$, 
one obtains the canonical form of a cubic equation,
\begin{equation}
\label{canon3}
\tau^3 + {\mathcal P}\tau + {\mathcal Q} =0\,,
\end{equation} 
whose analytical roots are extracted by the application of the 
Cardano-Tartaglia-del Ferro method. Noting that 
\begin{displaymath}
\label{peeque}
{\mathcal P}=\mu -\frac{1}{\lambda^2}
\left(1+\frac{\beta^2}{3}\right),\,  
{\mathcal Q}=-\frac{\beta}{3\lambda}
\left(\frac{2\beta^2+9}{9\lambda^2}+2\mu\right),
\end{displaymath} 
the discriminant of the cubic equation is 
\begin{equation}
\label{discri}
{\mathcal D}=\frac{\mathcal{Q}^2}{4}+\frac{\mathcal{P}^3}{27}\,.
\end{equation}
The sign of $\mathcal D$ is crucial here. If $\mathcal{D} >0$,
there will be only one real root of $\tau$ (the other two being
complex, which are known to occur in pairs). On the other hand, 
if $\mathcal{D} <0$, all the three roots of $\tau$ will be real. 
Knowing $\tau$, leads easily to knowing $t$, which, 
on using Eq.~(\ref{modeq}), gives the corresponding value of $N$. 
This is a turning point on the line $\dot{N}=0$. If there
are three real roots of $\tau$ (for $\mathcal{D} <0$), then
there will be three turning points of $N\left(t\right)$. 
If, however, 
there is only one real root of $\tau$ (for $\mathcal{D} >0$),
then there will be only one turning point of $N\left(t\right)$, 
the other two having been annihilated on the line, $\dot{N}=0$. 
\begin{figure}[floatfix]
\begin{center}
\includegraphics[scale=0.70, angle=0]{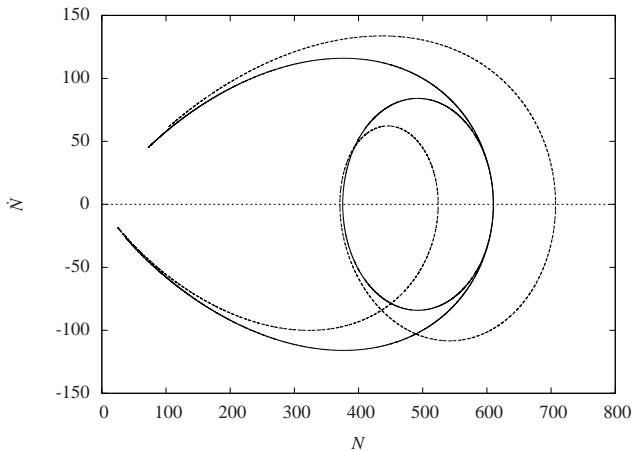}
\caption{\label{f5}\small{
Phase plot of the bimodal distribution of the traffic flow due
west. Both the curves are plotted for the values of $A$, $\mu$
and $\lambda$ used in Fig.~\ref{f1}. When $\beta =0$, the
resulting plot is shown by the continuous symmetric curve. When
$\beta =-0.09$ (as in Fig.~\ref{f1}), asymmetry sets in, as
shown by the dotted curve.
}}
\end{center}
\end{figure}
\begin{figure}[floatfix]
\begin{center}
\includegraphics[scale=0.70, angle=0]{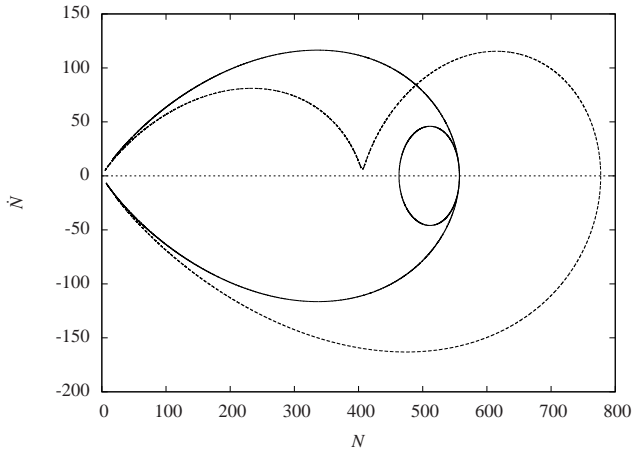}
\caption{\label{f6}\small{
Phase plot of the bimodal distribution of the traffic
flow due east. The plotting here follows the scheme of Fig.~\ref{f5}.
Both the curves are plotted for the values of $A$, $\mu$ and
$\lambda$ used in Fig.~\ref{f2}. The continuous curve shows the
symmetric case (for $\beta =0$), while the asymmetric dotted
curve is plotted for $\beta =0.24$ (as in Fig.~\ref{f2}). Since
this value of $\beta > \vert \beta_{\mathrm b} \vert$ (the
latter value is to be read
from Fig.~\ref{f4}), a bimodal-to-unimodal transition has already
taken place, and instead of a closed loop on the dotted curve, all
that remains is a cusp just above the line, $\dot{N}=0$.}}
\end{center}
\end{figure}

Evidently, a critical condition is reached when $\mathcal{D}=0$, 
i.e. when a bimodal-to-unimodal transition occurs. This might 
be viewed as a bifurcation in the bimodal system, for the 
value of $\beta =\beta_{\mathrm b}$. To understand how asymmetry 
in the bimodal distribution brings about this type of bifurcation
behaviour, one needs to follow how $\mathcal{D}$ depends 
on $\beta$. This variation has been plotted in Fig.~\ref{f4}, 
using the values of $\mu$ and $\lambda$ needed to calibrate 
the data of both the east and west-moving traffic. Considering 
$\vert \beta \vert$ (since $\beta <0$ for the west-going
traffic), a transition from a bimodal distribution to a 
unimodal distribution (or vice versa) can be seen to 
take place (when $\mathcal{D}=0$) for 
$\vert \beta_{\mathrm b} \vert \simeq 0.21$ 
in the case of the eastward 
traffic, and for $\vert \beta_{\mathrm b} \vert \simeq 0.46$ in the 
case of the westward traffic. 

Going back to Figs.~\ref{f1}~\&~\ref{f2}, one 
can see from the former plot, the value of $\vert \beta \vert$
needed to fit the data ($\vert \beta \vert =0.09$) is much less 
than what can cause the bimodal-unimodal bifurcation 
($\vert \beta_{\mathrm b} \vert \simeq 0.46$), while in the 
latter plot the reverse is true 
(here $\vert \beta_{\mathrm b} \vert \simeq 0.21 < \vert \beta \vert 
= 0.24$). The consequence of this is that, as opposed to 
Fig.~\ref{f1}, no conspicuous secondary peak in the 
bimodal distribution is seen in Fig.~\ref{f2}. The implications of 
both these conditions, as far as the dynamics on the $N$--$\dot{N}$ 
phase portrait is concerned, have been shown in 
Figs.~\ref{f5}~\&~\ref{f6}. These two plots show the dynamics
of bimodality for the west-moving and the east-moving traffic,
respectively. In both the cases, when $\beta =0$, there is a point
on the line $\dot{N}=0$, where the curve loops over itself. The
shifting of this point away from the aforesaid line, suggests
breaking of symmetry and asymmetric bimodality. For $\beta <0$
(westward traffic), this point shifts below the line, while for
$\beta >0$ (eastward traffic), the point shifts above. So to 
give a succinct description, it can be said that $\mu$ causes 
bimodality, and $\beta$ causes a breaking of symmetry. Taking 
these two aspects of the two parameters 
together, what results is an asymmetric bimodality. 
And changing the values of both these parameters can cause a 
bimodal-to-unimodal transition in the distribution, much 
in the likeness of a bifurcation. 

\section{Concluding Remarks}
\label{sec5}

The dynamics and the phase-portrait analysis, based on the model
equation for bimodal behaviour, have helped  
in unearthing some general properties of a bimodal distribution. 
The first is that the signature of bimodality in a phase-portrait
is the existence of a closed loop, and the second is the existence 
of a homoclinic trajectory, connecting a lone fixed point to 
itself in the phase portrait. Arguably, these are features which 
will hold true in other types of bimodal distribution as well. 
Although the primary emphasis of this study is to understand 
the generic mathematical features of bimodality, its practical
implications might help in designing innovative traffic 
control mechanisms, and afford a new perspective on addressing 
traffic-related problems.

The model developed here is based on a relatively uncomplicated
bi-directional flow of traffic along one axis only (in this case,
east to west
and vice versa). There is scope for refinement, when one considers 
the complications associated with the vast traffic networks of 
very big cities. A large vehicular traffic flow in a particular 
direction is the outcome of global decision patterns of commuters. 
In addition, traffic flows depend on routing possibilities in a 
traffic network, as well as on congestions that 
can occur in that network. In other words, the physics of 
vehicular traffic flows is associated with a complex 
dynamic competition between the demand for traffic 
space and its availability. The bimodal pattern studied here 
gives a simple one-dimensional perspective of this complex 
structure and its dynamics. 
With the aid of advanced modelling techniques, it should become 
possible to simulate and analyse these conditions, for which data 
may not be available, or for situations in which it may be 
difficult to perform
realistic experiments on large scales. All of these should go a 
long way in building a generalised model that could be applied
to traffic flows in an arbitrarily structured network.  

\begin{acknowledgments}
The authors thank R. Atre, A. Basu, J. K. Bhattacharjee 
and T. Naskar for some useful discussions. 
\end{acknowledgments}

\bibliography{bbt_mr2}

\end{document}